\documentclass[11pt]{article}
\usepackage{amssymb}
\usepackage{amsthm}
\usepackage{amsmath}
\usepackage{xspace}
\usepackage{url}
\usepackage{amsfonts}
\newcommand\skipit[1]{}

\newcommand\Bool{\mathbb B}

\title {\textbf{Virtual Evidence:\\
 A Constructive Semantics for Classical Logics }}

\skipit{ \title {\textbf{A Constructive Semantics for Classical Logics}}  }

\author {Robert L. Constable}

\date{ }

\begin{document}
\maketitle

\begin{abstract}

\noindent This article presents a computational semantics for classical logic using constructive type theory. Such  semantics seems \emph{prima facie} impossible because classical logic allows the \emph{Law of Excluded Middle} (LEM), not accepted in constructive logic since it does not have computational meaning. However, the apparently oracular powers expressed in the LEM, that for any proposition $P$ either it or its negation, $\sim P$, is true can also be explained in terms of constructive evidence that does not refer to ``oracles for truth." Types with \emph{virtual evidence} and the constructive impossibility of negative evidence provide sufficient \emph{semantic} grounds for classical truth and have a simple computational meaning. This idea is formalized using \emph{refinement types}, a concept of constructive type theory used since 1984 and explained here. A new axiom creating \emph{virtual evidence} fully retains the constructive meaning of the logical operators in classical contexts.  \\

\noindent A key step in this approach is motivated by observations of the twenty two old A. Kolmogorov
who showed in 1924/25 how to translate classical propositional logic into constructive propositional logic by using $\sim \sim P$ in place of $P$. His method is explained here. It was extended by Kuroda to first order logic and modified by G{\"o}del for number theory. That method and its modifications do not provide a new \emph{semantic account} of classical logic, rather they rely on syntactically translating classical propositions to constructive ones. The new idea proposed here is to define \emph{virtual constructive evidence} for \emph{classical propositions} using the \emph{refinement type} of computational type theory to specify the classical computational content. The following refinement type, $\{Unit |P \},$ is critical. Its only element is the unique element $\star$ of $Unit$, provided $P$ is known, otherwise it is the empty type. If $P$ is known by constructive evidence $p$, then the refinement type has $\star$ as its only \emph{explicit} element, and $p$ is hidden \emph{implicit evidence}. We imagine that $p$ has been ``squashed" to $\star,$ and we call $\{Unit |P \}$ ``squashed $P$." \\

\noindent The new axiom, $\sim \sim P \Rightarrow \{Unit | P \},$ creates virtual evidence for $P.$ That evidence can be used to construct virtual evidence in other refinement types by following the standard rules for them. There is constructive evidence for $\sim \sim P$ but only virtual evidence for $P.$  We abbreviate $\{Unit |P \}$ to $\{P\}$, so the new axiom is simply $\sim \sim P \Rightarrow \{P\}.$  This axiom, called here \emph{Classical Introduction}, is similar to the law of \emph{Double Negation Elimination} (DNE) favored by Kolmogorov to axiomatize classical logic as an extension of constructive logic. If there is only constructive evidence for $\sim \sim P,$ and none known for $P,$ then there is only virtual evidence for $\{P \}$ that is also hidden and can be unhidden only when proving classical propositions.  \emph{Virtual constructive evidence carries more information than is provided by standard classical semantics}. This new semantics integrates classical and constructive logic and is possible because constructive type theory expresses more kinds of evidence than classical set theory. \\


\end{abstract}

\flushleft
\setlength{\parskip}{0.05in}

\newpage

\section{Introduction}

Constructive logic is important in computer science for several reasons. One of them is that proof assistants for constructive type theory are widely used in formal methods, software development, and as programming assistants for dependently typed programming languages.\footnote{As a consequence, they can express refinement types, quotient types, intersection types, partial types, and rich recursive and co-recursive types.} Another is that constructive logics support the \emph{correct-by-construction} and \emph{secure-by-construction} programming methods. Furthermore, computational type theory appears to be an appropriate \emph{foundation for computer science} because computation is essential to the theory. The computational content of propositions and proofs is given by computable operations and discrete data. Also the origins of computer science as a discipline have deep roots in the investigations of the foundations of mathematics. Indeed, computer science has created technologies that draw on Hilbert's \emph{formalism}, Brouwer's \emph{intuitionism}, and Russell's \emph{type theory}. The central ideas of computability and unsolvability from Church, Turing, and Kleene lie at the heart of both computer science and constructive logic.

In this context it is important to understand more completely the relationship between classical logic and constructive logic.
A great deal is known about the relationship between these two kinds of logic from the
seminal work of Kolmogorov \cite{Kol24,Dra07}, Glivenko \cite{Gli29}, G{\"o}del \cite{God58,God65}, Kuroda \cite{Kur51}, Kleene \cite{Kle52}, Friedman \cite{Fri78} and others who will be mentioned here.\footnote{For an English translation of Kolmogorov's 1924 article see \cite{SBML67}, for very informative comments about it see Uspensky \cite{Usp92}. The cited G{\"o}del article is an English translation of a four page 1933 article. This approach might be abbreviated as the KGKG (Kolmogorov,Glivenko,Kuroda,G{\"o}del) approach, or $(KG)^2.$ } In addition results about typing programming language control operators \cite{Gri90,Mur91,Gir91} reveal interesting connections to classical logic, e.g. that some control operators have classical typing. These results (except for Kleene's) are based on \emph{proof theory} which is by nature concerned with details of specific proof systems and axioms. Proof theory ties the work to particular axioms, inference rules, and deduction systems. In contrast, the account given here is \emph{semantic}. It can be formalized in all proof systems of various styles, from the Hilbert style and the Gentzen styles \cite{Gen69} (Natural Deduction and Sequent Calculus -- either standard bottom up or top down as in LCF based systems \cite{GMW79}) to the Beth Semantic Tableaux \cite{Smu68,Fit69}.

Semantic approaches are less dependent on the details of axioms and inference rules, and they are used to
justify and explain formal presentations of the rules. The semantic approach was used by Allen \cite{All87a} to validate the rules of constructive type theory implemented by Nuprl. This method was formalized in Coq by Anand and Rahli \cite{AR14}.
Semantic methods typically require a richly expressive metatheory such as set theory
for classical logics and type theory for constructive ones. \emph{The
expressive richness of type theory with \emph{dependent types} is critical to the semantic nature of these new results and methods.} Semantic approaches have been used to justify a classical subset of the constructive type theory CTT14 implemented by Nuprl, (see Howe \cite{Howe96}), and a classical version of the calculus of inductive constructions implemented by Coq \cite{BYC04} is justified by proof theoretical results. It is not clear whether the Coq approach currently supports mixed mode assertions of the kind illustrated later in this article.

This article is an introduction to a semantic approach based on virtual evidence. We can achieve a complete understanding of this method in the case of propositional logic, proving that the new rule used here is consistent and complete for this classical logic. We have a good understanding for first-order logic as well where consistency is guaranteed and there are forms of completeness as well. For Peano Arithmetic (PA) the situation is more subtle with respect to Heyting arithemitc (HA), which can express undecidability and incompleteness.

For type theory as a whole, many interesting questions arise. For example, the new semantics supports a wide variety of \emph{mixed mode assertions}, blending constructive and classical, as illustrated later in the article. At this level of conceptual richness, interesting questions arise about integrating classical logic in a natural way. We can imagine results that would engage the mathematical community widely.

\subsection{Summary of the constructive semantics of classical logic}

In a sense, mathematics has been constructive since Euclid, and so called ``classical mathematics" is a 19th century trend. There are topics in modern mainstream mathematics that have always been constructive. Also, given a choice, mathematicians generally favor constructive proofs. Edwards \cite{Edw06} gives a modern look at this constructive line as does Henrici \cite{Hen88} and other articles too numerous to mention. Intuitionistic mathematics is an approach to constructive mathematics developed by L.E.J. Brouwer starting with his doctoral thesis in 1907 and developed over his lifetime in a series of deeply novel and fundamental mathematical results \cite{Bro75,vSt90,VAt04}.\footnote{Brouwer
called his work \emph{modern mathematics}, see \cite{VAt04}. Kolmogorov appears to have been the first logician to axiomatize Brouwer's ideas for \emph{intuitionistic} mathematics.  Brouwer is also considered one of the founders of topology and is well known for his classical fixed point theorems.} Moreover, Kolmogorov suggested a unification of intuitionistic and classical logics based on his \emph{double negation embedding} of classical logics into intuitionistic ones. Glivenko \cite{Gli29} proved they were equally expressive for propositional logic, and Kuroda \cite{Kur51} extended the result to first-order logic. Murthy \cite{Mur91} implemented this Kolmogorov embedding in Nuprl and used it in his automated formal transformation of a classical proof of Higman's Lemma into a constructive proof. Kolmogorov's approach would also allow mixing of classical and constructive logic. This is accomplished here as well based on the intuitive computational semantics we provide for classical logic.


In his article about Kolmogorov's work \cite{Dra07}, Drago explores the use of \emph{doubly negated sentences} (DNS) to give a constructive meaning to results in scientific papers. He shows that many scientific claims have the form of a DNS.\footnote{My colleague Mark Bickford and I have found similar applications in understanding statements about distributed systems \cite{BC08}.} This article provides a semantic explanation for these results using concepts from constructive type theory that were not available until the 1980's.

\paragraph{The Key Idea} The new idea presented here involves two steps. The \textbf{first step} is to hide (or suppress) the evidence for a proposition $P$ using the \emph{refinement type}. The classical meaning of $P$ is based on the constructive refinement type $\{Unit | P\}$ where $Unit$ is the type with precisely one element, $\star$. To know $\{Unit | P\}$ constructively, one needs to know evidence $p$ for $P$, but it is not included with the evidence type. Only $\star$ is an element of this type when it is non empty. The evidence $p$ is virtual in this setting. We also call this refinement type \emph{squashed P}. It is clear that there is evidence for $P \Rightarrow \{Unit | P \}$, so we know this implication constructively. However, we do not know $\{Unit | P \} \Rightarrow P$ in general because we are only assuming that there is evidence for $P$, and even if we had known evidence for $P$ and hidden it, to prove this proposition, we would need to reconstruct that evidence. Clearly we do not have evidence for $\{Unit | P ~\vee \sim P \} \Rightarrow (P ~\vee \sim P)$ for all propositions $P$.

\skipit{We will see below that the computational evidence for $P \Rightarrow \{Unit | P \}$ is a function that takes evidence for $P$ and produces $\star$ in $\{Unit | P \}$ and discards the evidence for $P$. To know that this function has this type requires that we have evidence for $P$ \emph{even though we don't save it with the type}. We don't carry evidence for $P$ with the type as a matter of efficiency since it might require a substantial amount of space that is not needed in computations. But this reason opens the possibility to define virtual evidence as we do next.}

Another way to explain the meaning of the refinement type $\{Unit | P\}$ is in terms of a general refinement type of the
form $\{x:A | B(x) \}$, which consists of those elements $a$ of type $A$ such that evidence $b$ for $B(a)$ is known. Thus an element belongs to $\{Unit | P\}$ if and only if it belongs to $Unit$ and evidence $p$ for $P$ is known. So if constructive evidence $p$ for $P$ is known, then $\star$ belongs to
$\{Unit | P\}.$ One can interpret this type as ``squashing" the evidence $p$ which might be complex, down to a single point, $\star.$  The constructive details are hidden, but this can only be done once evidence is known in the first place or built up from assumptions that have this classical character. This ``hiding operation" was introduced in ``Constructive Mathematics as a Programming Logic {I}: Some Principles of Theory" \cite{Con85} and used extensively since. This step does not take us beyond constructive logic, but it introduces the idea that there is a type that is designed to hide evidence.  Now we look at properties of this type in preparation for the second step.

If we know $\sim P$ constructively, then we know that $\{Unit | P\}$ is definitely empty -- there is nothing to hide.
When we know constructively that there is no evidence for $\sim P$, then we know that $\{Unit | \sim \sim P\}$ \emph{cannot be empty}. We also know that it can have only one possible unhidden element, namely $\star.$ Inspired by Kolmogorov's observation that the classical axiom for Double Negation Elimination, $\sim \sim P \Rightarrow P$, gives us full classical logic, we ponder whether we can make constructive sense of the idea that $\sim \sim P \Rightarrow \{Unit | P \}$. \emph{If we don't actually have evidence for $P$, so we can't  constructively inhabit the type} $\{Unit | P\}.$ To do so requires evidence for $P$, even though once we have that evidence, we hide it.\footnote{In proof assistants such as Nuprl and MetaPRL, the evidence can be found in the proof object, but it is not part of the refinement type.} We call this hidden evidence \emph{virtual} because we can use $\{Unit | P\}$ \emph{as if we had evidence for }$P$ when we are trying to prove other refinement types. That is how the refinement rules work.

For classical reasoning, constructive evidence for $P$ is not required. Classical reasoners are contemplating the possibility that there might be evidence available in ways yet unknown. Perhaps an
\emph{Omniscient Assistant} (OA) knows convincing evidence. Once a deeper understanding is achieved using virtual evidence, it might be possible to find constructive evidence. This has often happened in the past.

In light of the above observations, the \textbf{second step} is adding the following new axiom of \textbf{Classical Introduction (CI)}: $$\sim \sim P \Rightarrow \{Unit | P \}.$$

The required constructive evidence to know that $\star$ belongs to this type is knowing specific evidence $p$ for $P$.\footnote{We might also want to name this Constructive DNE.} We might not have such evidence since we only assume we know $\sim \sim P$. In particular, we do not have it when we consider $\sim \sim (P ~\vee \sim P) \Rightarrow \{Unit | (P ~\vee \sim P)\}$ even though we have constructive evidence for $\sim \sim (P ~\vee \sim P).$

Our proposal is to forego including evidence for $P$ when realizing the new axiom $\sim \sim P \Rightarrow \{Unit | P \}.$  \emph{This step agrees with the classical notion that evidence is not required for axioms}. It agrees with the constructive notion that axioms should be realizable in that the constant function whose value is $\star$ \emph{is an adequate realizer for the only evidence that must be visible}. The missing evidence is not carried along with the type, and when the type occurs in assumptions, only the virtual evidence is available. That virtual constructive evidence is all that is necessary for classical reasoning. The only explicit evidence needed for computation is $\star.$ Since constructively $\{Unit | P\}$ can not be empty, and since it can only have $\star$ as a member, this must be the member. \emph{So this axiom is constructive in the sense that we have not lost the required computational meaning of the type.}

To recapitulate, if there is constructive evidence for $\sim \sim P$, call it $nnp,$ then we can provide a function to give us the \emph{accessible evidence} for $\{Unit | P\}$, namely $\star$. That is all we are allowed to use
constructively from the type. The constant function whose value is $\star$ is the realizer for this axiom.  \emph{There is no loss of classical content} when we make this positive assertion that we have trivial evidence $\star$ for $P$.  Thus we can provide \emph{computational evidence} for the implication $\sim \sim P \Rightarrow \{Unit | P \}$, that is, evidence for $\sim \sim P \Rightarrow \{P\}.$ With this new axiom we can prove: $$\{Unit | P \} \Leftrightarrow  \sim \sim P.$$

The Classical Introduction axiom reveals the essential difference in evidence requirements between classical and constructive logic, and it allows us to conduct classical reasoning with refinement types in a way that can be explained constructively without assuming that we have an oracle that can provide enough evidence to justify the proposition constructively, yet \emph{can provide computational evidence that is sufficient for classical reasoning}. By mixing the classical and constructive modes, we allow a natural style of classical reasoning that supports extracting programs from proofs with some classical steps.

In this context, one way of understanding $\{P\}$ is that by proving this squashed proposition, one knows that it
is \emph{classically consistent}. In some cases there is constructive evidence in hand for $P$. In other cases the constructive status of $P$ might be entirely unknown, yet to be discovered or perhaps forgotten or only a partly scribbled note in the margin or perhaps evidence known only to an Omniscient Assistant that might be revealed one day or even discovered by a proof assistant that knows $\{P\}$ and is set the goal to keep looking for a proof of $P$. In any case, establishing $\{P\}$, raises the possibility of the stronger result $P$. Many constructive results have been discovered by exploiting this possibility.

What we also know is that the classical reasoning mode is semantically consistent with constructive reasoning. This will not be proved here. It can be established in the semantic model of CTT and could thus be formally proved in Coq.

\skipit{
We have two distinct choices, one to postulate that the type has this element -- that leads to classical logic with a \emph{(weak) constructive semantics}. The other choice is to avoid this kind of evidence because it does not lead to useful mental constructions or computations using $P$ itself. If we add the evidence, we can prove the squashed LEM, $\{P ~\vee \sim P\}$, from the one new axiom and the interesting constructive fact $\sim \sim(P ~\vee \sim P)$ telling us that \emph{there is no definite proposition which we can ever constructively prove to be unknowable}. So in reality, the binary choice we face is precisely, add the realizer or not. (This is a meta-level use of constructive $Act$ or $\sim Act$ with a clear \emph{binary choice}.)

The insight behind this new principle is that a constructive proof that a refinement type with \emph{at most one element} \emph{is known to be non empty} can be regarded as computationally realizable evidence that the type is inhabited by its unique element.
We can provide the computational content of this principle. In the idiom of classical logic, \emph{the corresponding classical proposition is true}. \emph{Even this (weak) constructive evidence carries more information than is provided by standard classical semantics.} This semantics for classical logic is possible because constructive type theory makes more logical distinctions than classical set theory, and thus it provides a richer semantic theory than the normal metatheory for classical logic. These distinctions allow new kinds of axioms. } 

\subsection{Markov's Principle}

In the CTT14 type theory, the identity propositions, $a = a ~in~ A,$ have $\star$ as a realizer.\footnote{Versions of constructive type theory are distinguished by writing CTT followed by the date they were implemented. Basically this theory has gotten progressively richer since its first release in 1984, CTT84 implemented by Nuprl. These are sometimes referred to collectively as the ``Nuprl type theory" since all versions are implemented by Nuprl, and they are upward compatible.} This is also the CTT84 method of asserting that $a$ has type $A,$ that is $a \in A.$ The equality rules extend this interpretation to all equality types, $a = b ~in~ A.$ We do not extend the classical modality to these equality types because we know from the work of Kopylov and Nogin \cite{KN01} that this will lead to a justification of \emph{Markov's Principle} (MP), which we do not regard as established by the arguments given here. This limitation of squashing is justified by the fact that we can in principle provide the equality evidence with the equality types, and there is explanatory value in doing that and using the squashing only as an efficiency step.

For people who would like to use classical reasoning to establish termination of computable functions, the methods of Kopylov and Nogin just mentioned can be seen as a natural extension of the virtual approach.\footnote{There are indeed researchers in the Coq community who are seeking such methods in a way that fits well with constructive type theory. They work well with Nuprl which has the refinement type, not yet included in the Coq type theory, the Calculus of Inductive Constructions (CIC) \cite{BYC04}.} We propose using the classical modality for this extension.

This constructive semantics for classical logic is studied here only for First-Order Logic (FOL). That it applies far more generally is clear -- indeed for all of constructive type theory. In that broad context, there are many interesting questions. The key principle involved has been used in constructive type theory for other purposes and is called \emph{``squashing" computational content}. It could be that we have collectively known all along that classical reasoning in the great preponderance of cases is not far removed from constructive reasoning. The apparently stronger classical assertions are the result of requiring weaker evidence for them than one expects for their constructive counterparts. Interestingly, the standard classical evidence given via set theory is inadequate for recovering the constructive substructure of classical logic. There is a sense in which type theory is a more suitable foundation for ``modern mathematics" than set theory. \\

\subsection{Comparing constructive and classical logic}

Many people believe that constructive logic is simply a subsystem of classical logic that does not use LEM.
This is misleading even for First-Order Logic (FOL) and incorrect unless the classical logical system uses all the logical operators, $\&,\vee, \Rightarrow, \forall, \exists$, and $False$ as primitives.\footnote{The negation of $P$, $\sim P,$ is defined as $P \Rightarrow False.$} The fundamental way in which this is the wrong view is that the constructive meaning of the logical operators is different from their classical meaning.  The constructive meaning is given in terms of \emph{evidence and computation} which are entirely missing from the semantics of classical logic. \emph{One of
the main benefits of the approach used here is that it retains the constructive meaning of the logical operators}, unlike in G{\"o}del's approach discussed later.

Constructive semantics expresses logical concepts in terms of mental constructions and basic computational operations of mathematics. In the case of \emph{minimal logic}, the semantics is completely unassailable and essentially universally accepted.
Minimal logic does not have the rule \emph{ex falso quodlibet}: from $False$ anything follows. That rule can be expressed as
$False \Rightarrow P.$ \footnote{It is said that Brouwer did not use this rule in his intuitionistic mathematics.} Moreover, for \emph{minimal first-order logic}, mFOL, we know that the standard proof rules are complete with respect to the intended interpretation of the logic, and that the completeness result is itself constructive \cite{CB14}.

On the other hand, when people are asked about why they believe classical logic, they typically talk about
the fact that the logical rules are ``obviously true."  Few people discuss ``Platonic reality." In the discussions with
my Cornell computer science colleagues that prompted this investigation, some of them said, ``well, I can't imagine that $P ~\vee \sim P$ could be false." In this view, they are entirely consistent with constructive mathematics. The problem with Platonic reality expressed in set theory is that naive mathematical truth is not at all clear and evident to everyone, and the formal accounts by Tarksi \cite{Tar56} and the standard textbooks that rely on either axiomatic or intuitive set theory \emph{already rely on classical logic} with LEM \cite{And86,Sho67,Smu68,CH07}! These theories do not make constructive sense even though FOL does.\footnote{Aczel \cite{Acz78} shows how to define a version of set theory in intuitionistic type theory, but not full ZFC set theory. Hickey implemented Aczel's approach in MetaPRL \cite{Hic01}.} Moreover, many basic questions about ``the Platonic reality of sets" have defied our efforts to clarify them. Perhaps we might have a better chance of seeing this reality in terms of types, as Church thought we might.\footnote{For example, it is easy to see in CTT that a constructive continuum hypothesis is false.}

In this article we will see a computational explanation of the classical laws of logic, and we can account for what people might mean when they believe that $(P ~\vee \sim P)$ is obviously true. What they often say when asked to defend this axiom is that for \emph{every well defined proposition} $P$, it is not possible to imagine any deeper explanation of LEM -- it is just ``obvious" and hence axiomatic. \emph{But it is far from universally obvious}. So one approach is to simply say the following: we cannot imagine that $\sim (P \vee \sim P)$ could ever be true since accepting that would require us to exhibit a proposition $P$ which we know to be unknowable. We cannot imagine how this would be done. \emph{Indeed, we constructively know that it cannot be done}.
Brouwer gave the following (unpublished) fully constructive argument for $\sim \sim (P ~\vee \sim P)$ which is reported on page 26 of Van Atten's small 2004 book \emph{On Brouwer} \cite{VAt04}:

``Can one ever prove of a proposition, that it can never be decided? No, because one would have to do so by reductio ad absurdum. So one would have to say: assume that the proposition has been decided in sense X, and from that deduce a contradiction. But then it would have been proved that not X is decided after all."

Here is a version of this informal argument that we create using constructive proof rules. To prove $\sim \sim (P ~\vee \sim P)$, we need to assume $\sim (P ~\vee \sim P)$ and prove a contradiction, e.g. prove $False.$ (This is Brouwer saying ``we need to use reduction ad absurdum.")  Our assumption is
$hyp1:(P ~\vee \sim P) \Rightarrow False.$ So we can prove $False$ if we can prove $(P ~\vee \sim P)$, so we take this as a goal, and chose to prove the $\sim P$ disjunct, the one on the right hand side of $\vee$.  Thus our subgoal is to assume $P$ and prove $False.$ (This is Brouwer saying, assume we decide to prove $P$ and discover a contradiction.) We now have two assumptions in play, $hyp1:(P ~\vee \sim P)$ and $hyp2:P$. We have two nested subgoals, both are to prove $False$. Now to prove the second (deepest) $False$ we use $hyp1$ a second time, again creating the subgoal to prove $(P ~\vee \sim P),$ but this time we chose to prove the left disjunct, $P.$ This is a good choice, because we have exactly this as our second hypothesis, so we are done. (Brouwer does not bother with these details, he notices that by proving $\sim P$ we have done the task we set ourselves and thus are done.)

We can present this argument symbolically after we define the sense of the constructive logical operators. This argument leads us to examine how one might justify LEM. We have seen that it is impossible that we cannot decide any specific proposition, say X. We cannot prove that X is unknowable. If that is the case, then for every X it must be possible to come to know it or its denial. This insight of Brouwer's is one of the most compelling reasons to believe
$\sim \sim (P ~\vee \sim P).$

We will see that this insight leads us to a plausible constructive explanation for $(P ~\vee \sim P)$. The explanations are in terms of how we come to believe mathematical
claims by finding evidence either for or against them. This evidence must lead to a \emph{mental experience} enabled by our ability to make mental constructions and carry out computations on representations of
mathematical objects.

\subsection{Consistency and completeness}

We can easily show that if $\{G\}$ is provable in iPC, then $G$ is provable in PC. This shows that the semantic approach is consistent. The proof is by induction on the (top down) sequent proof tree. The key steps are justifying the Classical Intro rule and the unfolding rule for the refinement type. We have discussed both of these already.

\textbf{iPC Consistency Theorem:} If $\vdash_{iPC} \{G\}$ then $\vdash_{PC} G.$

We can also show that the method is complete in the sense that we can prove any result of the classical propositional calculus.

\textbf{iPC Completeness Theorem:} If $\vdash_{PC} G$, then $\vdash_{\{iPC\}} \{G\}.$

This proof is very simple. We only need to replace the rule of double negation elimination, all of the other rules are
constructive. We do that by invoking the Classical Introduction rule.

It is also relatively easy to prove a consistency theorem for iFOL, but the issue of completeness is complicated by wanting to find
the most general fact to capture in a theorem. We cannot use the principle $$\sim \sim \forall x.(P(x) ~\vee \sim P(x)) \Rightarrow \{\forall x.(P(x) ~\vee \sim P(x))\}$$ because $\sim \sim \forall x.(P(x) ~\vee \sim P(x))$ is not constructively valid. It can easily be seen that it is not \emph{uniformly valid}, and thus by the completeness theorem for this concept, it is not provable in iFOL, see \cite{CB14}. We need to settle for justifying $\forall x.\{P(x) ~\vee \sim P(x) \}$, which is valid and seems sufficient. We do not pursue the completeness issue in this article.

\textbf{iFOL Consistency Theorem:} If $\vdash_{\{iFOL\}} \{G\}$ then $\vdash_{FOL} G.$

For arithmetic, the classical theory of Peano Arithmetic (PA) is inconsistent with constructive
arithmetic, Heyting Arithmetic (HA), because in HA there are \emph{unsolvable problems}, for instance, the halting of the i-th partial recursive function, say $\phi_i$ on its own index, $Halt(\phi_i(i))$ is undecidable. Kleene \cite{Kle52} discusses this issue.

\textbf{Classical Incompatibility Theorem:} $\vdash_{HA}~ \sim (\forall i: \mathbb{N}.(Halt(\phi_i(i)) ~\vee \sim Halt(\phi_i(i)) ),$ and $\vdash_{PA}~ \forall i: \mathbb{N}.( Halt(\phi_i(i)) ~\vee \sim Halt(\phi_i(i)) ).$

The obvious adjustment for both FOL and PA is to restrict the scope of the universal quantifier. The classical claim can be expressed as

\textbf{Classical Solvability Theorem:} $\vdash_{\{iPA \}}~ \forall n: \mathbb{N}.\{Halt(\phi_n(n)) ~\vee \sim Halt(\phi_n(n))\}.$

G{\"o}del provided a general translation of classical number theory into intuitionistic number theory by giving a translation of classical propositions into constructive ones that do not mention disjunction or the existential quantifier. His translation is given below where $P$ are atomic propositions, and $A$ and $B$ are arbitrary propositions of Peano Arithmetic.

\begin{enumerate}

\item Given atomic proposition, $P$ let $P^{o}$ be $P.$
\item $(A \Rightarrow B)^{o}$ is $(A^{o} \Rightarrow B^{o}).$
\item $(A \& B)^{o}$ is $(A^{o} \& B^{o}).$
\item $(A \vee B)^{o}$ is $\sim(\sim A^{o} \& \sim B^{o}).$
\item $(\sim A)^{o}$ is $\sim A^{o}.$
\item $(\forall x.A(x))^{o}$ is $\forall x.A^{o}(x).$
\item $(\exists x.A(x))^{o}$ is $\sim \forall x.\sim A^{o}(x).$

\end{enumerate}

Notice that the translated formulas do not express propositions using
either the constructive $\vee$ (or operator) nor the constructive existential quantifier, $\exists x.$
These are among the most interesting logical operators since they express
our ability to make decisions and construct objects. In classical logic,
we give up on those modes of expression and thus lose a great deal of
the computational meaning expressible in constructive logic. In our approach,
the constructive meaning of the logical operators is preserved.

Given a list of labeled formulas $H$ as hypotheses, let $H^{o}$ be
the list with each formula translated. We know this result.

\textbf{G{\"o}del Translation Theorem:} If $H \vdash G$, then $H^{o} \vdash G^{o}.$

\section{Propositions-as-types semantics}

In this section we briefly define the \emph{intended constructive interpretation} of the logical operators, using the propositions-as-types principle. We express the principle in constructive type theory. We begin with a brief historical sketch putting this semantics in context and relating it to other variants of the idea.

\subsection{Historical context}

The central concept in logic is usually taken to be \emph{propositions}, and the central questions are how do we understand them, i.e. grasp their \emph{sense}, and how can we \emph{know} them. The classical approach to knowledge is to define truth with respect to an ideal mathematical universe described in set theory. Tarski \cite{Tar56} is most associated with this idea, and it gives rise to \emph{model theory} \cite{CK73}, a branch of modern logic.

The constructive approach to truth is to understand how knowledge is represented in logic and how we formalize it. These are central questions in philosophy, and there are many approaches to clarifying them. We take an approach informed by a philosophical analysis provided by logicians, mathematicians and philosophers. Martin-L{\"o}f \cite{ML83} and Tait \cite{Tai83} provide compact summaries and Dummett \cite{Dum77} a longer one. There is a vast literature on the subject. The summary here is based on my enthusiasm for the idea that \emph{constructive mathematics is concerned with finding evidence that leads us to believe a proposition whose sense we grasp}. At one end of the spectrum of evidence is the notion of a formal proof, and idea developed in depth by Hilbert \cite{Hil67} and his school. At the other end is the notion of mental constructions and our intuitions about numbers and time studied by intuitionists such as Brouwer \cite{vSt90}.

The semantics of constructive/intuitionistic logic in the spirit of Brouwer can be given using
the \emph{propositions as types principle}, an idea with many variants and going by several other
names. It can be used for classical logic as well. It is employed here and defined below. Brouwer started developing this style of semantics informally by 1907 \cite{Bro75}. Kolmogorov axiomatized the propositional calculus in 1924 \cite{Kol24,Kol67}, and then it became possible to provide a semantics account of the axioms.  Heyting axiomatized first-order logic by 1935 \cite{Hey34}. Kleene provided a semantic account based on partial recursive functions and called it \emph{realizability} \cite{Kle45}. Curry \cite{CF58} understood the propositions as types semantics for propositional reasoning, and Howard \cite{How80} extended Curry's approach to give us the \emph{Curry-Howard isomorphism} \cite{SU06}. De Bruijn saw that propositions as types had a classical meaning as well \cite{deB70,NGDV94}, and the author employed it to provide a \emph{semantics of evidence} for classical logic \cite{Con85m}. Martin-L{\"o}f used the idea in his various type theories \cite{ML98}.

The author and his colleagues used this constructive semantics to develop a programming logic/language \cite{BC85} and a constructive type theory (CTT84) implemented by Nuprl \cite{book-full}.  That work led to one of the slogans that summarizes the idea, namely \emph{proofs-as-programs} and \emph{programs-as-proofs}. Cordell Green \cite{Gre69} also recognized the value of this idea early on and developed software for \emph{correct-by-construction} programming \cite{SG96}. The constructive type theory has steadily evolved to the current version, CTT14, which has been formalized by Anand and Rahli \cite{AR14} in the calculus of inductive constructions (CIC) implemented by Coq \cite{BYC04}.
\footnote{CIC is a related constructive type theory, but it does not support the refinement type used for our classical semantics. There is likely a way to accomplish the same ends in the CIC constructive type theory.}

The key idea is that the meaning of a proposition, say $P$, is given by a type whose elements
can be understood as \emph{evidence} for knowing $P$. It must be evidence that people and machines can
produce, check, manipulate, communicate, and thus people can also experience. If expression $P$ has the syntactic form of a proposition, then we denote by $[P]$ the \emph{type of evidence for} $P$. Next we examine the key concepts for understanding first-order propositions. The semantic prerequisite for knowing that a concept is a proposition is knowing what constitutes \emph{evidence for believing it}. Each clause below explains that evidence precisely for the standard logical operators.

\subsection{The meaning of propositions}

The syntax of propositions will be the same as presented in classical logic and already briefly discussed above. The classical meaning of propositions is given in terms of the truth values in the type of Booleans $\Bool$, say $tt$ and $ff$. This makes it especially easy to say what atomic propositions $True$ and $False$ mean classically, namely they denote their corresponding truth values. It is also easy to say what a propositional variable such as $P$, $Q$, $R$, etc. mean. Their meaning is given by an environment that maps these variables into the Booleans, $\Bool$.

The constructive meaning of propositions is quite different. We have already seen that $False$ is defined as the empty type,
and $True$ is identified with a singleton type such as $Unit$. What about compound propositions?  Their meaning is given by other types. To understand them, we need to first understand types. We will rely on an intuitive account of types described as we go.
In addition one needs a basic understanding about computable functions when we discuss implication, $\Rightarrow$, and universal quantification, $\forall.$ This semantics also relies on understanding \emph{data} and how computers process it. Understanding data is critical when we look at concrete theories such as arithmetic, where the data are natural numbers. In this realm the difference between constructive logic and classical logic is pronounced, and the two views are in conflict. Nevertheless, we can integrate them with this semantics.

\begin{itemize}

\item \textbf{Atomic propositions}: If a \emph{propositional expression} has no internal structure, expressed only
by simple constants, as in $True$, $False$ then it is \emph{atomic}. We cannot use its structure to tell us what it means. For our purposes, the only propositional \emph{constants} we single out are these two. We also assume an unbounded number of atomic propositions denoted by $A,B,C,P,Q,R$ and so forth. As we discuss specific areas of mathematics, we will see compound propositions such as $\forall n:\mathbb{N}.\exists m:\mathbb{N}.n < m$, and we might abbreviate them by a letter such as $S_0$ since their structure is atomic relative to the logical operators we are investigating.

We already see in the simple case of atomic expressions that we need to acquire certain general concepts for explaining how we understand propositions. Among those is the idea of a collection of evidence, called a \emph{type}. We assume this idea in the metatheory. In addition is the idea of a correspondence between syntax and its meaning given by the mapping from a propositional expression to a type, denoted by $[~~ ]$.

\item \textbf{Conjunction}: Perhaps the simplest compound proposition to understand is the conjunction,
in English, ``P and Q," and symbolically either $P \& Q$ or $P \wedge Q$. Here we will use the $\&$ sign for conjunction. If the evidence type for $P$ is $[P]$ and for $Q$ is $[Q]$, then the evidence type for
$P \& Q$ is the Cartesian product $[P] \times [Q]$ consisting of the ordered pairs $<p,q>$, where $p$ is in the type $[P]$, and $q$ is in the type $[Q]$. This is very intuitive.

To know a conjunction, we need evidence for each conjunct collected together and associated with the conjunction so that we know which evidence is assigned to which proposition. Notice that here we have used the concept of \emph{ordered pairs}. So our type theory must have the notion of a \emph{Cartesian product},
a standard name for the type of ordered pairs. This is a semantically primitive
concept that people must grasp intuitively and machines must implement in order to process evidence. We know that people can do this, and we built machines this way. Our primitive mental apparatus reveals this concept to be something that all people understand.\footnote{Brouwer analyzed the primitive concepts needed to understand mathematics. He believed that \emph{time} is the fundamental concept humans understand, and that the \emph{separation of time into two distinct moments} gives rise to the notion of ``twoity." We see that this idea is used here to understand conjunction.}

\item \textbf{Disjunction}: The next most complex logical construct is the
disjunction, ``P or Q," symbolically, $P \vee Q$. Evidence for this kind of proposition must tell us
two things: for which \emph{disjunct}, $P$ or $Q$, do we have evidence; and what is that evidence. To accomplish the first, we use a tag on the evidence.  It could be a tag such as ``left" for the $P$ disjunct and ``right" for $Q$. Instead we use notation that has become standard, $inl$ and $inr$ standing for ``inject into the left side" and ``inject into the right side."  The objects being injected are
the evidence for the proposition selected, say $inl(p)$ for $P$ and $inr(q)$ for $Q$.  The type of
this kind of evidence is called a \emph{disjoint union type}, and it is denoted $[P] + [Q].$\footnote{This notion is the dual of ``twoity," and the human mind grasps this duality as a general concept for how
twoity can be used, either to \emph{combine} or to \emph{separate}.}

\item \textbf{Implication}: Implication is associated with our ability to compute, to take evidence
in one form and \emph{convert it} to evidence in another form. This ability is one of our mental tools
for making \emph{constructions}. From this ability we come to understand an important constructive
type, the type of \emph{effective operations} taking evidence of type $S$ and converting it to evidence of type $T.$ We denote this type of all the effective operations from type $[S]$ to type $[T]$ as $[S] \rightarrow [T].$ We will use the \emph{lambda calculus} notation for effective operations.\footnote{This notation for operations is due to Church \cite{Chu40} who simplified the type theory of Russell and of Russell and Whitehead in \emph{Principia Mathematica} \cite{Rus08,WR25}.} But first we offer a simpler notation familiar from any mathematics course, using a function name such as $f$ and writing an equation to define it, such as $f(x) = x$ for the identity function, $f(x) = 0$ for a constant function, $add(<x,y>) = x+y$ for addition, and so forth.

Let $f$ denote a \emph{computable operation} from the type $[S]$ to the type $[T]$;  we
write $f:[S \rightarrow T]$ to say this succinctly.  We could also use the term ``function" for this $f$, and we will discuss terminology later in more detail. We stipulate that $f$ makes sense on inputs $s$ of type $S$ in that it computes to a specific value, denoted $f(s)$ of type $[T]$.

When we define specific functions, such as the identity function, $id$, that has the property that
$id(x) = x$, we will use the well known \emph{lambda notation}, $\lambda(x.t(x))$, to define an operation from type $[S]$ to type $[T].$ Given any evidence $s$ in type $[S]$, then the expression $t(s)$, where evidence $s$ is \emph{substituted} for the variable $x$ in the expression $t(x)$, computes to (or reduces to) evidence in type $[T].$ In this notation, the identity function is $\lambda(x.x)$.

If $P$ and $Q$ are propositions, then so is the implication $P \Rightarrow Q$. Evidence for implication
is an effective operation $\lambda(x.q(x))$ such that given any evidence $p$ of type $[P]$, $q(p)$ reduces to evidence of type $[Q].$

\item \textbf{Domains of Discourse and Types}: Most of our mathematical discourse
and our computer programs refer to particular kinds of data or elements. Logicians and philosophers call this the \emph{domain of discourse}, and use $D$ to denote it. Programming languages classify objects by
their types, and we think of $D$ as one of these ``data types."

In classical discourse, $D$ is usually any non-empty set. In constructive discourse, it is \emph{any type}, including possibly empty types. In this article, we will be primarily interested in settings for which $D$ is well known and standard. However, we can express the notion that a domain $D$ can have \emph{virtual elements} by using $\{D\}.$ We have already seen the unit types, $Unit$ and $Void.$  We will want to consider numerical domains such as the type of natural numbers, $\mathbb{N}$, integers, $\mathbb{Z}$, rational numbers, $\mathbb{Q}$, or real numbers, $\mathbb{R}.$

To keep matters simple for comparison to classical logic, we will assume that the domain of discourse, $D$, is non-empty and that we know an element $d$ of $D$. We will use $Void$ as discussed above and the meaning of $False.$

\skipit{The collection of these various types is called a \emph{Large Type}, and for our purposes, we will need only one of these, called $Type$ or synonymously, $Prop.$}

\item \textbf{Propositional Functions}: Once we have $Type$ and a domain of discourse, $D$ in $Type$,
our propositions are about elements of $D$. They will be
based on atomic \emph{propositional functions}, also called \emph{predicates} or \emph{relations}, which are \emph{computable} functions from $D$ into $Prop.$ For example, over the natural numbers $\mathbb{N}$ we might talk about \emph{equality}, $n = m$ or \emph{order} $n < m.$ We define these as proposition valued functions from the domain $D$ to $Prop.$ Since we treat propositions as types, these are also functions from $D$ to $Type.$

In constructive type theory, the way that we say that an element $d$ belongs to a type $D$ is to
assert $d = d ~in~ D$, abbreviated as $d \in D.$ The evidence for this claim is the trivial evidence $\star$ that we also use for other atomic propositions such as $True.$ We can adopt this approach because every type $D$ must come with an equality relation on it, defining what it means for two elements of $D$ to be equal.

\item \textbf{Existence}: Saying that something exists having certain properties is an important
logical utterance. We write it as $\exists x.P(x).$ The evidence for this assertion is a pair,
$<d,p>$ where $d \epsilon D$ and $p$ belongs to $[P(d)]$, i.e. $p \epsilon [P(d)].$
This makes sense because $\exists x.P(x)$ is well formed only when $P$ has type $D \rightarrow Prop.$
Thus we know that $P(d)$ belongs to $Prop$ since we have already established $d \epsilon D.$

We see a dependency among the parts of this proposition and in the evidence. The evidence for $P(d)$
\emph{depends on} $d$ and the type of $P.$ This is evidence for saying ``we can construct an element $d$ of $D,$ such that $P(x).$"

\item \textbf{Universality Claims}: The most expressive claims in first-order logic have the form that
there is evidence that some property $P$ or relation $R$ holds for all elements of a domain of discourse.
We write these as $\forall x.P(x).$ The evidence here is quite substantial. It consists of an effective operation $f$ that takes any element $d$ of $D$ and produces evidence in $[P(d)],$ and this operation also respects the equality relation on $D$ so that it behaves as a function.\footnote{In the CTT constructive type theory, we do not accept Church's Thesis, so the
  effective operations might go beyond the Turing computable functions.} That is, if $d_1 = d_2 ~in~ D,$
then $f(d_1) = f(d_2) ~in~ [P(d_1)].$ We already know that $P(d_1)$ is equal to $P(d_2)$ in $Prop$ because $P$ is a propositional function, thus it respects equality on $D.$

\end{itemize}

\subsection{Squashing computational content}

The constructive type theory that we use to provide a rigorous semantics includes the notion of
a \emph{subset type}, a.k.a. \emph{refinement type} of the form $\{x:T | P(x)\}$ where $T$ is a type and $P$ is a propositional function over the type $T$. This type was defined and explored in 1985 \cite{Con85} and adopted in the constructive type theory of the book \emph{Implementing Mathematics with the Nuprl Proof Development System} \cite{book-full} in 1986. Its elements are the members of the type $T$ that satisfy the proposition $P(x)$, but the evidence for the proposition is hidden; it is not accessible from the type. The type
$\{Unit | P\}$ has one element exactly when there is constructive evidence for the
proposition $P$. Let us agree that the element of $Unit$ is $\star$. Then the squashed type
$\{Unit | P \}$ has the element $\star$ when there is constructive evidence for the proposition, and it
is empty otherwise.

\textbf{Definition}: For a proposition $P$, let $\{P\}$ be the corresponding \emph{squashed
proposition} $\{Unit | P\}$ where $P$ has the constructive meaning as explained above.

It is trivial to extend the logical operators to squashed propositions, e.g. to consider these propositions:
$\{P\} \& \{Q\}, \{P\} \vee \{Q\}, \{P\} \Rightarrow \{Q\}, \exists x.\{P(x)\}, \forall x.\{P(x)\}.$

Notice that for $\{P ~\vee \sim P \}$, the squashed Law of Excluded Middle (LEM), the type will be constructively empty if the proposition $P$ is not decidable. On the other hand, the type $\{\sim \sim (P ~\vee \sim P)\}$ is inhabited for any $P$ by $\star$.
This is because we can prove the constructive fact $\sim \sim (P ~\vee \sim P)$. It has quite interesting
computational content which we will exhibit later. But this is expected for constructive logic, and all we have done is hide the computational content.

\subsection{Double negation and virtual constructive evidence}

Now we introduce a simple but powerful new notion associated with the classical rule of logic favored by
Kolmogorov and adopted by Kleene and others to express classical logic by adding one new law to constructive logic, the law of \emph{Double Negation Elimination} (DNE).\footnote{In Kleene's \emph{Introduction to Metamathematics }\cite{Kle52}, this law is Rule 8, and the Wisconsin logic students referred to it
as ``the dreaded Rule 8."}

\textbf{Double Negation Elimination:} $\sim \sim P \Rightarrow P$.

In the constructive logic of squashed propositions, we have argued that there is virtual evidence
for $\sim \sim P \Rightarrow \{P\}$, and the function $\lambda(f.\star)$, is a realizer.

\skipit{the constant function from
$(P \rightarrow Void) \rightarrow Void$ whose only element is $\lambda(x.x)$ since
$(P \rightarrow Void)$ is $Void$ if $P$ is inhabited. If $P$ is uninhabited, then so is
$\sim \sim P \Rightarrow \{P\}.$ } 

We have noticed that $\sim \sim (P ~\vee \sim P)$ is constructively true and thus
$\{\sim \sim (P \vee \sim P)\}$ is inhabited precisely by
one and only one element, $\star$. This means that the following proposition is virtually constructively
true.

\textbf{Theorem:} $\forall P:Prop. \{P ~\vee \sim P\}.$ \\
\textbf{Proof}: We know $\sim \sim \{P ~\vee \sim P\}$ and $\sim \sim \{P\} \Rightarrow \{P\}$. \textbf{Qed}

Ideas similar to what is discussed here were used by Kopylov and Nogin in their article ``Markov's Principle for Propositional Type Theory" \cite{KN01}. They give an account of Markov's Principle. They also suggest a connection to classical logic that comes from a classical metatheory. Their principle shows how to treat certain kinds of classical proofs as constructive proofs. The relevant paragraph from their 2001 paper is quoted here.  I was an advisor on this paper, and we did not realize at the time the full implications of this idea of a squash modality, as we called $\{P\}$, as applying to all classical logics and classical semantics. We especially did not see adding the principle for weak constructive semantics.

We can make the interpretation of Double Negation Elimination (DNE) virtually constructive because we use the virtual evidence only in proving other propositions with the squashed modality, only for proving $\{P\}.$ Kopylov and Nogin also allow this kind of reasoning for proving equality relations, and hence for proving the typing judgments of constructive type theory.  As they show, this allows them to assert that \emph{Markov's Principle} (MP) is actually constructive, but they do not frame their task this way. They frame it in the context of a classical meta theory. That extends the idea considerably beyond the approach
taken here, but they provide a plausible account of MP if we use the squashed modality. Using the approach of this article,
one could provide an even more plausible reason that MP might be considered constructively true since
there is no need for a classical metatheory.

The main novelty in the approach of this article is that we use the squash modality only in a limited way and with essentially
constructive semantics. Here is another interesting theorem.

\textbf{Theorem:} $\forall P:Prop. \{P \} \Rightarrow \sim \sim P.$
\textbf{Proof:} We assume $\{P \}$ and assume $\sim P$ with a goal of proving $False.$ \\
\textbf{Qed}

For this goal we can unhide the squashed $\{P \}.$
\skipit{
The squash operator can be considered as a modality. The propositional logic
equipped with this modality can express a principle that allows turning classical
proofs of squash–stable propositions into constructive ones. This principle is
valid in a standard type theory semantics if we consider it in the classical meta-
theory. Therefore this principle does not destroy the constructive nature of type
theory in the sense that we can always extract a witness term from a derivation.
It turns out that this principle implies Markov’s principle providing us a
propositional analog of Markov’s principle. It is rather surprising that such ana-
log exists because normally one needs quantifiers in order to formulate Markov’s
principle.

We also show an equivalent way of defining the same principle using a membership type instead of the squash
operator.

}

\subsection{Axioms for refinement types}

The rules for refinement types in CTT are quite simple and clear. Here is essentially the account given
for CTT84 that explains the rules in CTT02, which remain in CTT14. The idea is that to prove
$\{x:A | B(x)\}$ from a list of hypotheses $H$, we need to specify an element $a$ and prove in a subgoal
that it has type $A.$ We then need to prove that this $a$ satisfies the predicate $B(x).$  This is called
the \emph{introduction rule}.  The rule format here is in the top down sequent style where the subgoals of the rule come after the goal and are generated from the goal and the rule name.

The decomposition rule tells us how to use a refinement type as an hypothesis.  The idea is that if our goal is to prove the formula $G$ from the hypotheses $H, u:\{x:A | B(x)\}, H'$, where $H$ and $H'$ are lists of
assumptions, then we open the refinement type to have access to the element $x$ of $A$ that we have assumed to
exist, and we use it to prove the goal.  But we do not have access to the proof of $B(x)$ for this purpose unless the goal is squashed.

\begin{itemize}

\item \textbf{Refinement Introduction}

                $H~ \vdash ~ \{x:A | B(x)\} ~by~ a$ \\
                $H ~ \vdash a \in A $ \\
                $H ~ \vdash B(a)$ \\

\item \textbf{Refinement Decomposition Normal Goal}

            $H, u:\{x:A | B(x)\}, H' \vdash G ~by~ open(u;x.slot)$\\

                $H, x:A, H' \vdash G ~by~ g(x) ~for~slot$\\

\item \textbf{Refinement Decomposition Squashed Goal}

            $H, u:\{x:A | B(x)\}, H' \vdash \{z:G | R(z)\} ~by~ open(u;x,v.slot(x,v))$\\

                $H, x:A, v:B(x), H' \vdash G ~by~ g(x,v)$ \\
                $H, x:A, v:B(x), H' \vdash R(g(x,v)) ~by~ r(x,v)$ \\

\end{itemize}

\section{Classical Propositional Calculus with Constructive Semantics}

The propositional calculus, PC, is the normal starting point for the
study of formal logic. It provides the meaning and proof rules for
propositions whose structure can be analyzed using the standard
logical connectives such as \emph{and}, \emph{or}, \emph{implies},
\emph{not} -- symbolically $\&,\vee, \Rightarrow, \sim$. These are not
the only logical operators that we use in mathematics and computing.
There are operators such as \emph{exclusive-or},
\emph{conditional-and}, \emph{strong-and}, and so forth.

On the other hand, we typically use only two of these operators to
define all the others when we present PC. For example, authors typically
use $\&,\sim$ or $\vee,\sim$ or $\Rightarrow, \sim$, and we could in fact
use just one primitive such as the Sheffer stroke, $P | Q$ saying that
$P$ and $Q$ are \emph{incompatible}, i.e. not both are true.\footnote{Can we use the \emph{Sheffer stroke} $P | Q$ ($P$ and $Q$ are incompatible, i.e. at most one of them is true) or
the \emph{joint denial}, $P \downarrow Q$ (both $P$ and $Q$ are false) to axiomatize
PC with exactly one connective -- these are the only two single operator possibilities according to
Smullyan \cite{Smu68}, p.14 Exercise 5.} In the case of the intuitionistic propositional calculus (iPC),
the basic logic operators are not definable one from another, and we need the full complement,
$\&,\vee, \Rightarrow,\sim$ or $\&,\vee, \Rightarrow, False.$

Thus, one of the first ``curiosities" we face in this endeavor is that to define PC constructively, we
need more primitives than just one or two, we need at least five including the refinement type,
$\{Unit | P\}.$ On the other hand, once we have defined all of the standard classical logical
operators, we can prove the classical reduction to just one operator.

Even more curiously, in the constructive case, we find that there are other binary logical operators
that are necessary to express computational type constructors natural in programming, such as the \emph{intersection operator} $\cup$ and notions of \emph{parallel or}, $||$, which we will not discuss here.

It is an interesting question whether there is some finite
collection of ``logical connectives'' that will suffice for all we
want to say. But it is a fact that all logic books and implemented
logics use these basic operators. Therefore we start with them, in particular
with the logical operation that is constructively most interesting: \emph{implication}.

Implication is also used in these notes to define negation.  To say
``not P'' for a proposition P, we say that ``P implies false'',
symbolically, $P \Rightarrow False$. We will study this notion in our
first logical system, \emph{minimal logic}, by first including a constant with no special
properties, denoted $\perp$. Then when we are ready to introduce
negation, we will specify that $\perp$ shall be interpreted as $False$ and
negation, $\sim P$ shall be defined as $P \Rightarrow False$. We will
see many advantages to this approach, one of them being that we can
give a computational explanation to negation.

\subsection{Logical Formulas}

We will start with a subsystem
called the \emph{minimal implicational calculus}. It is included in
every version of Propositional Calculus. This calculus has precisely
one propositional constant, $\perp$, exactly one logical operator,
$\Rightarrow$, for implication, and an unbounded list of
\emph{propositional variables} whose official representation is
$var\{char\}$ where $char$ is a character string. For example,
variable could be $var\{X\}$, $var\{Y\}$, $var\{Unknown \}$, and so
forth.  Normally we will write these more simply as $A$, $B$, $C$,
$P$, $Q$, $R$, $X$, $Y$, etc. We sometimes denote this simple
calculus as $PC[\perp,\Rightarrow]$.

The class $Form$ of \emph{formulas} of this calculus consists exactly of the objects given by this inductive definition.
\begin{itemize}
\item the constant $\perp$,
\item variables $var\{char\}$ for some character string in place of \emph{char},
\item for $F$ and $G$ formulas, $imp\{F;G\}$ is a formula,
\item nothing else is a formula.
\end{itemize}

Some of our explanations are simpler if we also include the \emph{constructive disjunction},
$or\{F;G\}$ as a formula. All the concepts below can be trivially extended
by including this clause.

This definition determines a class of objects, $Form$, whose
elements we call \emph{formulas}. This is a typical example of an
\emph{inductive definition}. We say what the atomic elements are,
$\perp$ and $var\{char\}$ (in this case an unbounded number of
them); and we say how to build new elements using the constructor
$imp\{A;B\}$. We stipulate that nothing else is a formula.

We can systematically generate all of the objects of the class.
Here are some examples listed out: $\perp$, $var\{X\}$,
$imp\{\perp;\perp\}$,$imp\{var\{X\};var\{X\}\}$,
$imp\{var\{X\};\perp\}$.

It is common to display implications, e.g. $imp\{var\{X\};\perp\}$, in
a more readable way, such as $X \Rightarrow \perp$. We call this the
\emph{display form} of the official syntax for a term in the class
$Form$. The official syntax is clear because there is no question
about the structure of the term. The outer operator of every formula
is either $\perp$, $var$, or $imp$. Only the term whose outer
operator is $imp$ has subterms. Specifically if the term in question
is $imp\{F;G\}$ for $F$ and $G$ formulas, then $F$ and $G$ are the
left and right subterms (subformulas) respectively.

It is immediately clear from the definition of formulas that they
have a unique structure, a unique way to read them.  This is not
true for the display forms unless we disambiguate them, say using
parentheses. For instance, if we see $A \Rightarrow B \Rightarrow
C$, we do not know whether the formula is $imp\{A;imp\{B;C\}\}$ or
$imp\{imp\{A;B\};C\}$. Thus we require that the display form show
the structure with explicit parentheses, e.g. the first version
displays as $A \Rightarrow (B \Rightarrow C)$, the second version as
$(A \Rightarrow B) \Rightarrow C$.

In some presentations of propositions there is a convention that the
$\Rightarrow$ operator associates to the right, thus in these
presentations, $A \Rightarrow B \Rightarrow C$, means $A \Rightarrow
(B \Rightarrow C)$.

\skipit{ A key feature of any logical system is some mechanism for defining new concepts from the primitives.
We use the notion of a \emph{definition} or abstraction to create new concepts from the primitives.
For example, we define a new primitive atomic proposition $True$ as $\perp \Rightarrow \perp.$ If we do not \emph{expand the definition}, then we treat the new primitive as an expression with no subterms. If we do expand it, then we see its simple structure, with two subterms $\perp$.}  

\subsection{Operational Semantics}

The idea of implication is something we understand as part of the
naive informal logic we use in the everyday world.  We don't often
use the term ``implication'' in informal discussion.  Instead of
saying $A \Rightarrow B$ is an implication or that ``proposition A
implies proposition B,'' we say ``A implies B'' or we might say, ``if
A then B.'' Usually when we say that A implies B or B follows from
A, we have some reason that we think we know this.  If someone says
``if it's snowing the temperature must be below freezing'', they
most likely know a reason for this, perhaps a bit of physics or
something they have always experienced. Some implications are the
result of piecing together bits of knowledge that constitute
evidence for why we believe the proposition. As we are learning
basic arithmetic, we tend to memorize some atomic justifications for
simple facts, e.g. $-5 \times -5$ is $25$ remembering that ``a minus
times a minus is a plus'' or some such rule.

We might know a name for a  reasoning step of the form $A
\Rightarrow B$ implies $\sim B \Rightarrow \sim A$ where we think,
``if $A$ implies $B$, then if not $B$, then not $A$.''  We will
study this law of logic, called the \emph{contrapositive}. Other
forms of inference were given names by the ancient logicians. These
names served as ``evidence'' for various assertions. Even some of
the logical fallacies had names such as ``affirming the consequent''
which is the claim that because we know $A \Rightarrow B$, and we
know $B$, we must know $A$.

In general, when we make assertions such as ``I know A'' or ``A is
true'' or just A, we have in mind some reason that we believe A to
be the case, we can provide \emph{evidence} for our claim.  Perhaps
for some simple assertions, such as $0=0$, we say: this is just what
equality means, or it is trivial to see this by comparing the symbols
on each side of the equality sign, they are identical.  If we see
$0=(0 \times 1)$, then we might need to say a bit more, such as,
$0\times 1$ computes to $0$ so this is the same as $0=0$. In general
what we know here is that if we discover that some expression $exp$
computes to $0$, then we know $0=exp$.

What is the reason that we know $A \Rightarrow A$ in general and
$\perp \Rightarrow \perp$ in particular? To answer this we need to
know what implication, $imp$ means. What does $A \Rightarrow B$ mean
in general? To know $A \Rightarrow B$, we need to reason from the
\emph{assumption} of $A$ to the \emph{conclusion} $B$, or from the
\emph{antecedent} to the \emph{consequent}. Thinking in terms of
evidence we see intuitively and naively that we know the implication
holds if we have a method of transforming evidence for the
assumption $A$ into evidence for the conclusion $B$, as in the case
of transforming the information that $exp$ evaluates to 0 into
evidence that 0 equals $exp$.

In general, we know $A \Rightarrow B$ at least when we have a
method that converts hypothetical evidence $x$ for $A$ into evidence
$b(x)$ for $B$.  We actually need to know the method and be able to
apply it. So let's give that transformation a name. We call it an
\emph{operation}, $f$, that transforms $x$ into $f(x)$ in such a way
that we know $f(x)$ is evidence for $B$ as long as we are given
real evidence for $x$, say $a$ for $A$, if there is any. Notice, this
transformation $f$ should work on hypothetical evidence. If this is
part of the way we really think and reason, then that operation $f$
must be effective in the sense that we can actually carry it out,
either ``in our heads'' or by doing some explicit computation, e.g.
as in reducing an expression $exp$ to $0$. Here are some very simple
examples.

We know $(A \Rightarrow (B \Rightarrow C)) \Rightarrow B \Rightarrow
(A \Rightarrow C)$. We can see this if we assume that $f$ is
evidence for $A \Rightarrow (B \Rightarrow C)$, then if $x$ is
evidence for $B$, then given $y$ as hypothetical evidence for $A$,
we see that $f(y)$ is evidence for $B \Rightarrow C$ and $f(y)(x)$
is evidence for $C$. We will see later how to put all this together
into a single function term.

We can now see why $\perp \Rightarrow \perp$ is  a proposition we
know. We reason that no matter what specific proposition $\perp$ is,
if we are given hypothetical evidence $x$ for the assumption, then
the transformation that produces evidence for the conclusion is the
identity transformation, $f(x)=x$. We have exactly the evidence
needed to know the implication.

We will see shortly that it will be very convenient to have compact
names for specific operations such as the identity transformation.
One notation will be $I$ for ``Identity''.  We also use the
functional notation from programming languages for this operation,
$\lambda(x.x)$.

We will start with this \emph{operational explanation} of
implication as one meaning that we definitely use in mathematics and
other areas of precise thinking, as in science, to understand
implication and its role in providing evidence or information that
causes us to accept an assertion because we understand it and have
the right kind of evidence for it. Let us see how this explanation
accounts for many of the laws typically expressed in the
Propositional Calculus such as these:

\begin{itemize}
\item $A \Rightarrow (B \Rightarrow A)$
\item $(A \Rightarrow B) \Rightarrow ((A \Rightarrow (B \Rightarrow C)) \Rightarrow (A \Rightarrow C))$
\item $(A \Rightarrow B) \Rightarrow ((B \Rightarrow C) \Rightarrow (A \Rightarrow C)).$
\end{itemize}

Using our notation for operations, we see that
$\lambda(x.\lambda(y.x))$ is evidence for the first proposition. It
is an operation that takes evidence $x$ for $A$ and returns an
operation $\lambda(y.x)$ which takes evidence $y$ for $B$ and
returns evidence for $A$, namely $x$. This is a very clean and
elegant explanation of how we can come to know an implication
without having specific evidence for any of the constituent
propositions, $A$, $B$, or $C$.  This illustrates how logic is
capable of working with hypothetical information and transforming
it.

The operational explanation of

$$(A \Rightarrow B) \Rightarrow ((A \Rightarrow (B \Rightarrow C)) \Rightarrow (A \Rightarrow C))$$

is the following operation

$$\lambda(f.\lambda(g.\lambda(x.g(x)(f(x))))).$$

What we have here is that $f$ is hypothetical evidence for the
assumption $A \Rightarrow B$, $g$ is hypothetical evidence for
$(A \Rightarrow (B \Rightarrow C))$, and $x$ is hypothetical evidence
for $A$. So we can see that $g(x)$ is evidence for $(B \Rightarrow
C)$ and $f(x)$ is evidence for $B$. Therefore by reasoning about the
composition of the operations $f$ and $g$, we see that $g(x)(f(x))$
is evidence for $C$ as required to have evidence for $A \Rightarrow
C$.  If we imagine being given concrete evidence for $f$ and $g$,
say $f_0$ and $g_0$, then $\lambda(x.g_0(f_0(x)))$ is indeed an
operation that takes evidence for $A$ into evidence for $C$.

We leave it as an exercise to find evidence for the last example.

These examples show us that in order to manipulate terms for
evidence, we should carefully develop the rules for applying
operations. We also recognize that these operations derive their
meaning from knowing the nature of the evidence they are
transforming.

To understand the evidence semantics, we see that the evidence for a
proposition $A$ can be thought of as data of a specific kind. Thus we
borrow a concept from one of the earliest investigations of logic,
the work of Bertrand Russell from 1908 \cite{Rus08,Rus08a}. He
defined a type as the range of significance of a propositional
function. Based on this idea applied to our operational meaning of
the logical operations from L.E.J. Brouwer in the same time period
\cite{Bro13,Bro75,vSt90}, we use the term \emph{type} to denote the
collection of terms that can be treated as evidence for a
proposition $A$.

Shortly we will identify the notion of proposition with its
semantics, with the type of all the evidence for it in a systematic
computational interpretation of the kind we just used to explain
implication.

As we progress in our understanding of logic, we will make more and more precise
the requirements on a type theory that make it a sufficient basis for the semantics
of the various logics studied in the literature on mathematical logic.
For the Propositional Calculus we will see that we need a type theory that treats functions as objects and thus for any two types, say $T_1$ and $T_2$, there will be the type of all effective operations taking input from $T_1$ into output in $T_2$.  This is the function type, denoted $T_1 \rightarrow T_2$, a standard type not only in mathematics but in modern programming languages. However, to call it a function type in the sense of mathematics, we need to define an equality relation on the types, usually denoted $x =_{T} y$ for equality on type $T$ or also $ x=y ~in~ T$. We will require that the evidence operations respect the equality, and when they do we call these operations \emph{functions}.\\

\noindent \textbf{Definition}: Given two types $T_1$ and $T_2$, the type of all effective operations from elements of $T_1$ into elements of $T_2$ is denoted $Op\{T_1;T_2\}$ and is called the type of \emph{effective operations} from type $T_1$ into type $T_2$. If the operations respect equality on the types, they are called \emph{effective functions} from $T_1$ into $T_2$, denoted $T_1 \rightarrow T_2$.\\

\noindent \textbf{Notation}: We denote elements of $Op\{T_1;T_2\}$ and $T_1 \rightarrow T_2$ by $\lambda(x.b(x))$ where if $t$ is an element of type $T_1$, then we know that $b(t)$ is an element of type $T_2$. If $f$ belongs to this type, then we denote its application to an element $t$ as $f(t)$ which we regard as the display form of the underlying primitive application term $ap(f;t)$.\\

\noindent \textbf{Reduction rule}: Given an operation $\lambda(x.b(x))$, the rule for reducing an application $ap(\lambda(x.b(x);t))$ regardless of the types, is to substitute the term $t$ for $x$ in $b(x)$ to obtain $b(t)$. For the moment we assume that the term $t$ is a data element and has no variables in it such as $x$.  This is an overly simple assumption as we will see later, and there are reasons for substituting terms $t$ which have free variables in them. When we do that we must be very careful not to confuse the occurrences of variables because that can change the meaning of the operation.  Also in the more detailed account below, we will not always indicate the body of the operation as a term $b(x)$ where we indicate the argument $x$ of the function explicitly. Instead we will use a notation for substitution such as $b[t/x]$ meaning to substitute the term $t$ into the expression $b$ for all free occurrences of the variable $x$.

\subsection{False Propositions and Negation}

A simple extension of the minimal implication logic allows us to
reason about negation.  We first consider the notion of a false
proposition in the setting of evidence semantics.  Let us specialize
the constant $\perp$ to the atomic proposition $False$. A false
proposition is one for which there is no evidence, so it's evidence
type should be empty. Let $Void$ be the empty type,
then $[False]$ = $Void$.

We denote this calculus as $PC[False,\Rightarrow]$.

Once we have the $False$ proposition, we can explain what it means that there is no evidence for a
proposition $P,$ by saying there is evidence for $P \Rightarrow False$. That evidence would have
the form $\lambda(x.p(x))$ where we know that $p(x)$ is evidence for $False$ under the assumption
that $x$ is evidence for $P.$
If there were evidence for $P$, then $\lambda(x.p(x))$ would convert it to evidence for $False$.
But since there is no evidence for $False$, this means there cannot be evidence for $P$ either.
Let us define the negation of a proposition this way.\\

\noindent \textbf{Definition}: $\sim P$ means $P \Rightarrow False$.\\

We just used a very important bit of logical reasoning in our analysis of $\sim P$.
We said that if some proposition $P$ implies the proposition $Q$, then if we know $\sim Q$,
we must know $\sim P$.  This is called the \emph{contrapositive} claim. We can already express
this in the new calculus with negation, we assert that $(P \Rightarrow Q) \Rightarrow (\sim Q \Rightarrow \sim P)$,
and we can build the evidence for this as follows. Let $pq$ be hypothetical evidence for $(P \Rightarrow Q)$,
let $nq$ be hypothetical evidence for $\sim Q;$ this is an operation to provide evidence for $Q \Rightarrow False$.
Now we need evidence for $P \Rightarrow False$ which is an operation from $P$ to $False$.  We build it from the other operations as:\\

$$\lambda(pq.\lambda(nq.\lambda(p.nq(pq(p))))).$$

This evidence expression captures the kind of informal logic that we used in analyzing the nature of negation. The empty type and the proposition $False$ are useful more generally as well.

A subtle point about this small calculus arises in thinking about what it should mean to assume $False$. What follows from this assumption? We surely have $False \Rightarrow False$ since we have this for any constant including $\perp$. We also have $False \Rightarrow True$ using the evidence $\lambda(x.\lambda(y.y))$, recalling that $True$ in this setting is $False \Rightarrow False$.

Do we have $False \Rightarrow P$ for any proposition $P$? There can be no evidence for $False$, so if we consider evidence of the form $\lambda(x.x)$ can this be evidence for $False \Rightarrow P$?  It is plausible to say that under the assumption that there is evidence for $False$, we can prove anything, including that $x$ is evidence for $P$. In ancient logic this was the ``\emph{ex falso quod libet}'' rule that claimed from false anything follows. This simple question has been studied and pondered a great deal, and one working conclusion is that it seems that no great harm is done adopting this rule and using something simple as evidence for $P$. We will return to this topic later, and in the meanwhile allow a rule adopted in Constructive Type Theory (CTT) and Intuitionistic Type Theory (ITT) that allows the following evidence term for this proposition, $\lambda(x.any(x))$.\\

\noindent \textbf{Special Evidence}: $\lambda(x.any(x))$ is evidence for $False \Rightarrow P$.\\

We can now prove $\sim P \Rightarrow (P \Rightarrow Q)$.

Here are some facts that we know about negation:

\begin{itemize}
\item $P \Rightarrow \sim \sim P$
\item $(P \Rightarrow Q) \Rightarrow (\sim Q \Rightarrow \sim P)$
\item $(P \Rightarrow \sim P) \Rightarrow (P \Rightarrow Q).$
\end{itemize}

Exercise: Is there evidence for $(\sim (P \Rightarrow Q)) \Rightarrow (P \Rightarrow \sim Q)$?

\subsection{Constructive semantics for classical PC}

Now we define the semantics of the classical propositional calculus, PC, using the constructive
operators. We need an interpretation or \emph{model}, $\mathcal{M},$ that assigns to the atomic propositions, $A,$
an atomic type, $\mathcal{M}(A).$ For compound propositions $P$, $\mathcal{M}(P)$ is the type assigned according the constructive semantics given above, the constructive meaning of $P$ according to
the propositions as types semantics. This definition is precise as it stands, but for increased
clarity we spell out the separate clauses of the definition as a review of the constructive semantics.

\begin{itemize}
\item if $P$ is atomic, then $\{P\} ~=~ \{Unit | \mathcal{M}(P)\},$
\item $\{P \& Q\} ~=~ \{Unit | \mathcal{M}(P) \times \mathcal{M}(Q)\},$
\item $\{P \vee Q\} ~=~ \{Unit | \mathcal{M}(P) + \mathcal{M}(Q)\},$
\item $\{P \Rightarrow Q\} ~=~ \{Unit | \mathcal{M}(P) \rightarrow \mathcal{M}(Q)\},$
\item $\{\sim P\} ~=~ \{Unit | \mathcal{M}(P) \rightarrow Void \}.$
\end{itemize}


\textbf{Definition}: We say that a formula $P$ of PC is \emph{classically valid} if and only if $\{P\}$ is
an inhabited type, thus, $\star$ belongs to $\{P\}$.

This semantics is clearly constructive, being defined in a version of constructive type theory. Using this semantics, we can validate the typical axioms of PC and prove that the inference rules preserve this notion of validity.

\textbf{Theorem 1}: All the standard axioms given for PC are valid under constructive evidence semantics for PC, and all the standard inference rules preserve validity. Therefore all provable formulas of PC are constructively valid.

The only interesting case is the Law of Excluded Middle (LEM), and we have discussed this extensively above. We discuss it a bit further below. This theorem requires the new \emph{Classical Introduction} axiom of extended constructive type theory, that $\sim \sim P \Rightarrow \{P\}$.

We also call this semantics \emph{virtually constructive evidence semantics} because the constructive meaning of
the logical operators is hidden or squashed or weak (from double negation), and is exposed only when proving squashed propositions. This restriction is what keeps the theory constructive.

The core of this new reasoning is the proof of $\{P ~\vee \sim P\}$, and that depends critically on the constructive proof of $\sim \sim (P ~\vee \sim P).$ This theorem tells us that we cannot ever construct any proposition $P$ that we \emph{know to be unknowable}. This is quite a nice insight that cannot be expressed so succinctly in classical logic. The constructive evidence for $\sim \sim (P ~\vee \sim P)$ turns out to be very interesting. In part, this is the case because it does not depend on using $False$ to express negation. It is a fact that we can prove without the law \emph{ex falso quod libet}, that from $False$ one can prove anything.

The following realizer shows that the formula $((P \vee (P \Rightarrow A)) \Rightarrow A) \Rightarrow A$ is constructively true.

$$ \lambda(h.h(inr(\lambda(p.h(inl(p) ))))).$$

This is a realizer for the type $$((P \vee (P \Rightarrow A)) \Rightarrow A) \Rightarrow A.$$

From this typing we see that the function $h$ has the type $((P \vee (P \Rightarrow A)) \Rightarrow A).$
It is also interesting that when we use arbitrary propositions such as $A$, we can see that the argument
is correct for $((P \vee (P \Rightarrow A_1)) \Rightarrow A_2) \Rightarrow A_2$ as long as we also know that
$A_2 \Rightarrow A_1.$ This means that if $A_1$ is $Void$, then $A_2$ must be $Void$ as well.

When $A = False$ this expresses the Law of Excluded Middle. Here we see that this is a more general fact. Friedman \cite{Fri78} was the first to significantly exploit this insight. 

We discuss other aspects of this constructive semantics in the next section that deal with first-order
quantifiers.

\textbf{Theorem 2}: All constructively valid formulas of PC are provable.

This result is constructively true since we can prove the completeness theorem constructively, there are several sources for this,
e.g. \cite{Cal98}.

\subsection{Kolmogorov embedding}

Kolmogorov \cite{Kol24} proposed and studied the following embedding of classical PC into the intuitionistic version, iPC.

\begin{itemize}

\item $K(P) \rightarrow \sim \sim P$
\item $K(P \& Q) \rightarrow \sim \sim(K(P) \& K(Q))$
\item $K(P \vee Q) \rightarrow \sim \sim (K(P) \vee K(Q))$
\item $K(P \Rightarrow Q) \rightarrow \sim \sim (K(P) \Rightarrow K(Q)$
\item $K(False) \rightarrow False$\footnote{This captures that $\sim \sim False$ is $False$.}

\end{itemize}

\textbf{Kolmogorov's Theorem}: If $P$ is provable in classical PC, then
 we can effectively transform any classical proof to a
 proof of $K(P)$ in intuitionistic PC.

 Applying the Kolmogorov transformation to a formula essentially doubles its
 size. For example, $$K( \sim(P \& Q) \Rightarrow ~ False )$$
 results in the formula
  $$(\sim \sim (\sim \sim (\sim \sim P \& \sim \sim Q) \Rightarrow False))$$
but in practice we could define classical versions of the logical
operators, say $\&^{'}$, $\vee^{'}$, $\Rightarrow^{'}$ and write the formulas this
way $(P \&^{'} Q) \Rightarrow^{'} False.$ Although this preserves the size of
the original formula, the indirect meaning provided by the translation is
quite difficult to grasp intuitively. By contrast, the semantic explanation
we provide here is intuitive in that it explicitly shows exactly what constructive
information is available and what is not.

\section{First-Order Logic}

The situation for First-Order Logic (FOL) is similar to that for PC. We can define $\{F\}$ for FOL formulas using
propositions as types and refinement types. The results go over without much difficulty using a standard constructive semantics for intuitionistic FOL (abbreviated iFOL). One can see the details of this logic in \emph{Constructivism in Mathematics, An Introduction} \cite{TvD88} by Troelstra and van Dalen.

\subsection{Constructive semantics for quantifiers}

It is quite clear how to extend the ideas from PC to FOL. Here is the definition of the classical modality for quantified formulas, we use the same ``squashing" mechanism as for PC.

\begin{itemize}
\item $\mathcal{M}( \{\forall x.P(x) \}) = \{Unit | \forall x.\mathcal{M}(P(x))\},$
\item $\mathcal{M} (\{\exists x.P(x) \}) = \{Unit | \exists x. \mathcal{M} (P(x))\}.$
\end{itemize}

The interesting point here is that for the classical proposition, $\forall x.(P(x) ~\vee \sim P(x)),$ we cannot establish $\{\forall x.(P(x) ~\vee \sim P(x))\}$ by proving $\sim \sim \forall x.(P(x) ~\vee \sim P(x))$ since this double negation is not constructively provable as we can see by using the notion of \emph{uniform validity} \cite{CB14} or Kleene's realizability semantics.

But all we need to establish is $\forall x.\{(P(x) ~\vee \sim P(x))\},$ which we can prove. In this form, the proposition does not claim that we can expose the decision made for each element of the domain, we simply use the fact that for any such element $d$ of the domain of discourse $D$, we know $\{P(d) ~\vee \sim P(d)\}.$

Kleene \cite{Kle52} in \emph{Introduction to Meta-Mathematics}, Theorem 63 on page 511 shows that the formula
$\sim \sim \forall x.(A(x) ~\vee \sim A(x))$ is not realizable for Heyting Arithmetic (constructive Peano Arithmetic) because $\sim \forall x.(A(x) ~\vee \sim A(x))$ is realizable.  This theorem arises from proving in HA that we cannot solve the Halting Problem for partial recursive functions.  So now we see that there are indeed unsolvable problems of this sort, not specific $P$ or $\sim P$ questions for a definite assertion $P$, but rather a claim about a class of problems $Halt(i).$ We know from results of Turing and G{\"o}del that such a class of problems can be unsolvable, hence the defining formulas are not constructively provable.

In the FOL case and for Peano Arithmetic (PA), we can seek to find the weakest use of the squash modality to preserve computational meaning. It is clear that if we apply the modality to the subformulas as well as at the outer level, we will have a result such as
$\forall x.\{P(x) \vee \sim P(x)\}.$

\textbf{Theorem 3}: All the standard axioms given for FOL are valid under constructive evidence semantics in which all subformulas are squashed. Moreover, all the standard inference rules preserve validity. Therefore all provable formulas of FOL are valid in the weak constructive semantics.

\section{Mixed Mode Logics}

It is possible in this context to blend constructive and classical logics and rely on constructive type
theory to provide the semantics. A similar mixed mode can be defined with respect to the Kolmogorov embedding. The semantics given here can provide a simple explanation for these mixed mode assertions. For example, we might ask these kinds of questions.

$$(P \Rightarrow Q) \Rightarrow \{P \Rightarrow Q\}?$$

This one is obvious, but what about this?

$$(\{P\} \Rightarrow \{Q\}) \Rightarrow \{P \Rightarrow Q\}.$$

What about these formulas?

$$\{P \& Q\} \Rightarrow \{P \} \& \{ Q \}?$$

$$\{P \} \& \{ Q \} \Rightarrow \{P \&  Q \}?$$

It interesting to mix the constructive and classical semantics. One of the most natural ways to do this
is to use our constructive semantics for classical propositions. For example, consider the conjecture,
$\sim \forall x.\{P(x) ~\vee \sim P(x) \}.$ What does this mean when $D$ is $\mathbb{N}$ and $P(i)$ is
the assertion $Halt(i),$ saying that program $\phi_i$ with numerical index $i$ halts on input $i$, e.g.
$Halt(i)$?  We know $\sim \forall x.(Halt(x) ~\vee \sim Halt(x))$ is consistent with Heyting Arithmetic (HA), so we do not know
$\sim \sim \forall x:\mathbb{N}.(Halt(x) ~\vee \sim Halt(x))$ is not provable in iFOL. We know $\forall x:\mathbb{N}.\{Halt(x) ~\vee \sim Halt(x)\}.$ We know in general that $\forall x.\{P(x) ~\vee \sim P(x)\}.$


\textbf{Acknowledgements} I would like to thank Juris Hartmanis for recent stimulating conversations on the topic of
constructive versus classical logics and thank my colleagues in the PRL research group, Mark Bickford, Abhishek Anand, and Vincent Rahli, Anne Trostle, and Sarah Sernaker for insightful comments on the way to present these ideas and helping me explore them in Nuprl.


\bibliographystyle{plain}
\bibliography{anne-rc-new}

\end{document}